# Impossibility of Spontaneously Rotating Time Crystals: A No-Go Theorem

Patrick Bruno*

*European Synchrotron Radiation Facility, BP 220, F-38043 Grenoble Cedex 9, France*

I present arguments indicating the impossibility of spontaneously rotating "quantum time crystals," as recently proposed by Frank Wilczek. In particular, I prove a "no-go theorem," rigorously ruling out the possibility of spontaneous ground-state (or thermal equilibrium) rotation for a broad class of systems.

We are familiar with the idea that setting something into motion entails some energy cost. Yet, this generally accepted paradigm has been challenged in a Letter [1] by Frank Wilczek who proposed the existence of a new state of matter, "quantum time crystals", defined as systems which, in their quantum mechanical ground state, display a time-dependent behavior (periodic oscillation) of some physical observable [2]. Wilczek's proposal has stimulated both a considerable interest [3–5] and a vivid controversial debate [6–9]. The proposal is based upon a model consisting of particles moving on an Aharonov-Bohm (AB) ring threaded by a magnetic flux, with attractive interaction. From the observation of the well-known facts that (i) in absence of coupling, a nonzero AB flux gives rise to a stationary (time-independent) ground-state current and that (ii) for zero AB flux, a sufficiently strong interaction induces a density modulation (soliton) on the ring, Wilczek then went on to argue that the combined effect of the AB flux and interaction would result in a spontaneous breaking of the time-translation invariance, with a persistent rotation of the soliton in the ground state [1]. However, in a recent Comment [6], I pointed out that Wilczek's rotating soliton is not the correct ground state of the model and that a static solution with a lower energy can be found; acknowledging this point, Wilczek nevertheless speculated in a Reply [7] that some other models could display a "quantum time-crystal" ground-state motion. A further general objection raised in Ref. [6], which was left unanswered in Ref. [7], is the fact that a system displaying a rotational motion in its ground state would be able to radiate energy (e.g., electromagnetic waves), which would conflict with the principle of energy conservation. In order to settle these puzzling questions, I give in the present Letter a general argument for the impossibility of spontaneously rotating quantum time crystals, based upon a "no-go theorem," strictly ruling out spontaneous ground-state (or thermal equilibrium) rotation for a broad class of systems with arbitrary composition and interactions.

The collective rotational dynamics of interacting systems has been investigated in detail in the past, in particular, in the context of rotating nuclei [10, 11] or, more recently, ultracold atomic gases [12, 13]; however, the presence of an AB flux, which breaks time-reversal invariance, is likely to modify substantially the physical behavior of the system and compels us to reexamine the problem in this new context. On the other hand, the effect of the AB flux is at the heart of Kohn's theory of the insulating state [14]. Kohn considered a system in a torus geometry threaded by an AB flux (or, equivalently, with twisted periodic boundary condition) and concluded that the hallmark of an insulator is its complete insensitivity to the AB flux, as a result of the localization of the many-body wave function. This consideration already suggests that the realization of a quantum time crystal by setting a ring-shaped Wigner crystal (which is known to be an insulator) into spontaneous ground-state rotation by threading the ring with an AB flux, as proposed in Ref. [3], is a hopelessly doomed endeavor.

A system can be meaningfully said to be in rotational motion only if it breaks rotational invariance in the first place. For finite systems subject to a potential with rotational symmetry, the ground state (like any energy eigenstate) will always be rotationally invariant [15]; thus, the rotational symmetry has to be explicitly broken by the external potential. The breaking of the rotational symmetry may also occur spontaneously, as a result of interactions, in the thermodynamic limit, as in Wilczek's model [1]. The correct treatment of the thermodynamic limit needs great care; a general method to this aim has been presented by Bogoliubov [16]: it consists of calculating physical quantities for finite values of the particle number $N$ and symmetry-breaking potential $V$, and then taking the limit $N \to \infty$ first, and the limit $V \to 0$ next [17]. In view of the above considerations, it appears that the problem amounts to studying the ground-state energy of the system $E_0^{(\Omega)}$, as seen from the static frame, for finite $N$ and in presence of an external symmetry-breaking potential rotating at angular velocity $\Omega$ [18] (this procedure is essentially equivalent to the cranking model of rotating nuclei [19]), and eventually taking the thermodynamic limit according to Bogoliubov's prescription. The system is a quantum time crystal if, and only if, the minimum of $E_0^{(\Omega)}$ is obtained for some nonzero value of $\Omega$. Such a behavior would take place, in particular, if the term linear in $\Omega$, in the series expansion of $E_0^{(\Omega)}$ in powers of $\Omega$, is nonvanishing. Since both the AB flux and the external potential rotation break the time-reversal symmetry of the Hamiltonian, terms of odd powers of $\Omega$ are generally allowed for noninteger AB flux, so that Wilczek's idea seems to be *a priori* plausible. Yet, as we shall show below, the term linear in $\Omega$ is always exactly zero. Furthermore, I prove that $E_0^{(\Omega)} > E_0^{(0)}$, for $\Omega \neq 0$. The latter inequality is a no-go theorem which strictly prohibits the existence of spontaneously rotating time crystals as speculated by Wilczek. These results are also generalized to the situation of thermal equilibrium at nonzero temperature.

Let us now move on to the proof of the no-go theorem. Let us consider an assembly of $N$ particles of masses $m_i$ moving on a one-dimensional AB ring (of radius $R$) threaded by a magnetic flux, described by the following Hamiltonian:

$$\hat{H}_{\boldsymbol{\phi}}^{(\Omega)}(\boldsymbol{\theta}, t) = \sum_{i=1}^{N} \left[ \frac{\hbar^2 (\hat{l}_i - \phi_i)^2}{2\, m_i R^2} + V_i(\theta_i - \Omega t) \right] + \sum_{i<j} U_{ij}(\theta_i - \theta_j), \quad (1)$$

where $\theta_i$ is the angular coordinate of particle $i$, and $\hat{l}_i \equiv -i\partial_{\theta_i}$ is the corresponding (dimensionless) angular momentum operator. The particles may be fermions or bosons (or any mixture of bosons and fermions) or discernable particles. The AB flux may be either a true magnetic flux or an effective gauge flux [20] due, for instance, to adiabatic spin tracking [21], to trap rotation [22], or to coherent level transitions [23]. For the sake of generality, we allow the flux, external potential, and interparticle coupling to take different values for each particle; this would be the case, for instance, if the various particles are located on physically different rings. The dimensionless number $\phi_i$ is the flux (in units of flux quanta) experienced by particle $i$, $V_i(\theta_i - \Omega t)$ is the external potential experienced by particle $i$ (which we set into rotation at angular velocity $\Omega$, as explained above) and $U_{ij}(\theta_i - \theta_j)$ is the coupling between particles $i$ and $j$. We use vector notations, such as $\boldsymbol{\theta} \equiv (\theta_1, \theta_2, \ldots, \theta_N)$ and $\boldsymbol{\phi} \equiv (\phi_1, \phi_2, \ldots, \phi_N)$. The generic model given by Eq. (1) encompasses not only the models considered in Refs. [1] and [3] but also the Fermi-Hubbard and Bose-Hubbard models for ultracold atomic gases on optical lattices [24, 25].

Let $\psi_{\boldsymbol{\phi},n}^{(0)}(\boldsymbol{\theta}, t) = \varphi_{\boldsymbol{\phi},n}^{(0)}(\boldsymbol{\theta})\, e^{-iE_{\boldsymbol{\phi},n}^{(0)} t/\hbar}$ be the $n$th many-body eigenstate of the static Hamiltonian $\hat{H}_{\boldsymbol{\phi}}^{(0)}(\boldsymbol{\theta})$, with energy eigenvalue $E_{\boldsymbol{\phi},n}^{(0)} \equiv \langle \psi_{\boldsymbol{\phi},n}^{(0)} | \hat{H}_{\boldsymbol{\phi}}^{(0)} | \psi_{\boldsymbol{\phi},n}^{(0)} \rangle$ (throughout the Letter, wave functions are normalized to 1). For nonzero $\Omega$, let $\psi_{\boldsymbol{\phi},n}^{(\Omega)}(\boldsymbol{\theta}, t)$ be the rotatory eigenstates [18] of the time-dependent Schrödinger equation $i\hbar \partial_t \psi(\boldsymbol{\theta}, t) = \hat{H}_{\boldsymbol{\phi}}^{(\Omega)}(\boldsymbol{\theta}, t) \psi(\boldsymbol{\theta}, t)$, with time-independent energy in the static frame $E_{\boldsymbol{\phi},n}^{(\Omega)} \equiv \langle \psi_{\boldsymbol{\phi},n}^{(\Omega)} | \hat{H}_{\boldsymbol{\phi}}^{(\Omega)} | \psi_{\boldsymbol{\phi},n}^{(\Omega)} \rangle$.

The transformation to the rotating frame is achieved by the change of variables $(\theta_i, t) \to (\theta_i', t') \equiv (\theta_i - \Omega t, t)$ (throughout this Letter, the primes indicate quantities expressed in the rotating frame), which yields $\partial_{\theta_i} = \partial_{\theta_i'}$ and $\partial_t = \partial_{t'} - \Omega \partial_{\theta_i'}$, so that the Schrödinger equation in the rotating frame becomes $i\hbar \partial_{t'} \psi'(\boldsymbol{\theta}', t') = \hat{H}_{\boldsymbol{\phi}}^{'(\Omega)}(\boldsymbol{\theta}', t') \psi'(\boldsymbol{\theta}', t')$, with $\hat{H}_{\boldsymbol{\phi}}^{'(\Omega)}(\boldsymbol{\theta}') = \hat{H}_{\boldsymbol{\phi}}^{(0)}(\boldsymbol{\theta}') - \Omega \hat{L}_z$. This is a classic textbook result [26]; here, $\hat{L}_z = \hbar \sum_i \hat{l}_i = \hat{L}_z'$ is the total angular momentum operator and is the same in the rotating frame as in the static frame. Simple manipulations then yield

$$\hat{H}_{\boldsymbol{\phi}}^{'(\Omega)}(\boldsymbol{\theta}') = \hat{H}_{\boldsymbol{\phi}+\Omega\boldsymbol{\mu}}^{(0)}(\boldsymbol{\theta}') - \hbar\Omega\Phi_{\text{tot}} - \frac{I_{\text{cl}}\Omega^2}{2}, \quad (2)$$

where $\Phi_{\text{tot}} = \sum_i \phi_i$ is the total flux seen by all the particles, and $I_{\text{cl}} = \sum_i m_i R^2$ is the classical moment of inertia of the system; here and below, we use the shorthand notations $\mu_i \equiv m_i R^2/\hbar$ and $\boldsymbol{\mu} \equiv (\mu_1, \mu_2, \ldots, \mu_N)$. Thus, the rotatory eigenstates and eigenvalues, expressed in the rotating frame, are, respectively,

$$\psi_{\boldsymbol{\phi},n}^{'(\Omega)}(\boldsymbol{\theta}', t') = \varphi_{\boldsymbol{\phi}+\Omega\boldsymbol{\mu},n}^{(0)}(\boldsymbol{\theta}') e^{-iE_{\boldsymbol{\phi},n}^{'(\Omega)} t'/\hbar} \quad (3)$$

and

$$E_{\boldsymbol{\phi},n}^{'(\Omega)} = E_{\boldsymbol{\phi}+\Omega\boldsymbol{\mu},n}^{(0)} - \hbar\Omega\Phi_{\text{tot}} - \frac{I_{\text{cl}}\Omega^2}{2}. \quad (4)$$

Their expressions in the static frame are thus, respectively,

$$\psi_{\boldsymbol{\phi},n}^{(\Omega)}(\boldsymbol{\theta}, t) = \varphi_{\boldsymbol{\phi}+\Omega\boldsymbol{\mu},n}^{(0)}(\boldsymbol{\theta} - \Omega t) e^{-iE_{\boldsymbol{\phi},n}^{'(\Omega)} t/\hbar} \quad (5)$$

and

$$\begin{aligned} E_{\boldsymbol{\phi},n}^{(\Omega)} &\equiv \langle \psi_{\boldsymbol{\phi},n}^{(\Omega)} | \hat{H}_{\boldsymbol{\phi}}^{(\Omega)} | \psi_{\boldsymbol{\phi},n}^{(\Omega)} \rangle \\ &= \langle \varphi_{\boldsymbol{\phi}+\Omega\boldsymbol{\mu},n}^{(0)} | \hat{H}_{\boldsymbol{\phi}}^{(0)} | \varphi_{\boldsymbol{\phi}+\Omega\boldsymbol{\mu},n}^{(0)} \rangle \\ &= E_{\boldsymbol{\phi}+\Omega\boldsymbol{\mu},n}^{(0)} - \langle \varphi_{\boldsymbol{\phi}+\Omega\boldsymbol{\mu},n}^{(0)} | \Delta\hat{H}_{\boldsymbol{\phi}}^{(\Omega)} | \varphi_{\boldsymbol{\phi}+\Omega\boldsymbol{\mu},n}^{(0)} \rangle, \end{aligned}$$

with

$$\begin{aligned} \Delta\hat{H}_{\boldsymbol{\phi}}^{(\Omega)} &\equiv \hat{H}_{\boldsymbol{\phi}+\Omega\boldsymbol{\mu}}^{(0)} - \hat{H}_{\boldsymbol{\phi}}^{(0)} \\ &= \hbar \sum_i \frac{\left[\hat{l}_i - (\phi_i + \mu_i\Omega)\right]^2 - (\hat{l}_i - \phi_i)^2}{2\mu_i} \quad (6) \\ &= \Omega \frac{d\hat{H}_{\boldsymbol{\phi}+\Omega\boldsymbol{\mu}}^{(0)}}{d\Omega} - \frac{I_{\text{cl}}\Omega^2}{2}. \quad (7) \end{aligned}$$

We eventually obtain

$$E_{\boldsymbol{\phi},n}^{(\Omega)} = E_{\boldsymbol{\phi}+\Omega\boldsymbol{\mu},n}^{(0)} - \Omega \frac{dE_{\boldsymbol{\phi}+\Omega\boldsymbol{\mu},n}^{(0)}}{d\Omega} + \frac{I_{\text{cl}}\Omega^2}{2}. \quad (8)$$

Obviously, the term linear in $\Omega$, in the expansion of $E_{\boldsymbol{\phi},n}^{(\Omega)}$, vanishes, as announced above, and one has

$$E_{\boldsymbol{\phi},n}^{(\Omega)} - E_{\boldsymbol{\phi},n}^{(0)} = \frac{I_{\boldsymbol{\phi},n}\Omega^2}{2} + O(\Omega^3) \quad (9)$$

where

$$I_{\phi,n} = I_{cl} - \left.\frac{d^2 E^{(0)}_{\phi+\Omega\mu,n}}{d\Omega^2}\right|_{\Omega=0} \quad (10)$$

is the moment of inertia for the $n$th energy level of the system. The second term, in the above equation, is the quantum correction, which gives rise to the phenomenon of nonclassical rotational inertia in coherent quantum systems such as nuclei [11] or superfluids [27]. It is also the central ingredient in Kohn's theory of the insulating state [14], who relates it to $\lim_{\omega\to 0}\omega\text{Im}\sigma(\omega)$. To obtain the moment of inertia, we expand the rotatory energy eigenvalue, expressed in the rotating frame, in powers of $\Omega$ and use second order perturbation theory; this yields $E'^{(\Omega)}_{\phi,n} = E^{(0)}_{\phi,n} - \Omega L_{\phi,n} - \frac{I_{\phi,n}\Omega^2}{2} + O(\Omega^3)$, with

$$L_{\phi,n} = \hbar\Phi_{tot} - \left.\frac{dE^{(0)}_{\phi+\Omega\mu,n}}{d\Omega}\right|_{\Omega=0}, \quad (11)$$

and

$$I_{\phi,n} = 2\sum_{m\neq n}\frac{|L_{\phi,n,m}|^2}{E^{(0)}_{\phi,m} - E^{(0)}_{\phi,n}}, \quad (12)$$

where $L_{\phi,n,m} \equiv \langle\varphi^0_{\phi,n}|\hat{L}_z|\varphi^0_{\phi,m}\rangle$ and $L_{\phi,n,} \equiv \delta_{n,m}L_{\phi,n,m}$. Obviously, for a ground state which breaks rotational symmetry, $I_{\phi,0} > 0$, and $I_{\phi,0} = 0$ otherwise. The latter expression for the moment of inertia generalizes the result obtained in absence of magnetic field for nuclei [10] or ultracold atomic gases [12]. For an excited state, the moment of inertia $I_{\phi,n}$ ($n > 0$) may be negative.

To generalize these results to the case of a system in thermal equilibrium, we first note that, since the reference frame in which the Hamiltonian is time independent is the rotating frame, the statistics of level population is controlled by the energy levels in the rotating frame $E'^{(\Omega)}_{\phi,n}$; thus, the free energy in the rotating frame $\mathcal{F}'^{(\Omega)}_{\phi,\beta}$, at temperature $k_BT \equiv \beta^{-1}$, is given by

$$e^{-\beta\mathcal{F}'^{(\Omega)}_{\phi,\beta}} \equiv \text{Tr}\left(e^{-\beta\left(\hat{H}^{(0)}_\phi - \Omega\hat{L}_z\right)}\right) = \sum_n e^{-\beta E'^{(\Omega)}_{\phi,n}}. \quad (13)$$

To obtain the expression of the free energy in the static frame $\mathcal{F}^{(\Omega)}_{\phi,\beta}$, we should carefully pay attention to the fact that $e^{-\beta\mathcal{F}^{(\Omega)}_{\phi,\beta}} \neq \sum_n e^{-\beta E^{(\Omega)}_{\phi,n}}$. Instead, one gets

$$\mathcal{F}^{(\Omega)}_{\phi,\beta} = \mathcal{F}'^{(\Omega)}_{\phi,\beta} + \sum_n\left[e^{-\beta(E'^{(\Omega)}_{\phi,n} - \mathcal{F}'^{(\Omega)}_{\phi,\beta})}\left(E^{(\Omega)}_{\phi,n} - E'^{(\Omega)}_{\phi,n}\right)\right]. \quad (14)$$

Inserting in the above equations the expressions given by Eqs. (4, 8), one obtains [28]

$$\mathcal{F}'^{(\Omega)}_{\phi,\beta} = \mathcal{F}^{(0)}_{\phi,\beta} - \Omega\mathcal{L}_{\phi,\beta} - \frac{\tilde{\mathcal{I}}_{\phi,\beta}\Omega^2}{2} + O(\Omega^3), \quad (15)$$

$$\mathcal{F}^{(\Omega)}_{\phi,\beta} = \mathcal{F}^{(0)}_{\phi,\beta} + \frac{\tilde{\mathcal{I}}_{\phi,\beta}\Omega^2}{2} + O(\Omega^3), \quad (16)$$

with

$$\mathcal{L}_{\phi,\beta} \equiv \langle\hat{L}_z\rangle_{\phi,\beta}, \quad (17)$$

$$\tilde{\mathcal{I}}_{\phi,\beta} \equiv \mathcal{I}_{\phi,\beta} + \beta\left(\langle\hat{\Lambda}^2_\phi\rangle_{\phi,\beta} - \langle\hat{\Lambda}_\phi\rangle^2_{\phi,\beta}\right), \quad (18)$$

$$\mathcal{I}_{\phi,\beta} \equiv \sum_{n\neq m}\left(\frac{e^{-\beta E^{(0)}_{\phi,n}} - e^{-\beta E^{(0)}_{\phi,m}}}{e^{-\beta\mathcal{F}^{(0)}_{\phi,\beta}}\left(E^{(0)}_{\phi,m} - E^{(0)}_{\phi,n}\right)}|L_{\phi,n,m}|^2\right), \quad (19)$$

where $\hat{\Lambda}_\phi \equiv \sum_n|\varphi^0_{\phi,n}\rangle L_{\phi,n}\langle\varphi^0_{\phi,n}|$ is the diagonal part of the angular momentum, and $\langle\hat{A}\rangle_{\phi,\beta} \equiv \text{Tr}\left[e^{-\beta(\hat{H}^{(0)}_\phi - \mathcal{F}^{(0)}_{\phi,\beta})}\hat{A}\right]$. In the expression of the moment of inertia $\tilde{\mathcal{I}}_{\phi,\beta}$, Eq. (18), the second term arises from the flux dependence of level populations; it vanishes for zero flux or zero temperature and was therefore absent in the results published earlier for systems with time-reversal invariance such as ultracold atomic gases [12]. Obviously, $\tilde{\mathcal{I}}_{\phi,\beta} > 0$ for a system in which rotational symmetry is broken at equilibrium, and $\tilde{\mathcal{I}}_{\phi,\beta} = 0$ otherwise.

Having shown that setting the system into rotation always increases the ground-state energy (or the free energy, for $T > 0$) in the limit $\Omega \to 0$, we finally show that the same holds for any finite value of $\Omega$. From Eqs. (8, 10), and from analogous relations for the free energy [28], we obtain, respectively,

$$E^{(\Omega)}_{\phi,0} - E^{(0)}_{\phi,0} = \int_0^\Omega d\Omega'\,\Omega' I_{\phi+\Omega'\mu,0} > 0, \quad (20)$$

$$\mathcal{F}^{(\Omega)}_{\phi,\beta} - \mathcal{F}^{(0)}_{\phi,\beta} = \int_0^\Omega d\Omega'\,\Omega'\tilde{\mathcal{I}}_{\phi+\Omega'\mu,\beta} > 0, \quad (21)$$

where the inequalities follow from the fact that $I_{\phi+\Omega'\mu,0} > 0$ and $\tilde{\mathcal{I}}_{\phi+\Omega'\mu,\beta} > 0$, respectively, in at least some finite range of $\Omega'$ for a system with broken rotational invariance at rotation frequency $\Omega$. The latter results hold for any values of particle number $N$ and symmetry-breaking potential $V$. Thus, when taking the thermodynamic limit according to Bogoliubov's prescription [16, 29], and assuming the occurrence of spontaneous breaking of rotational symmetry, we obtain for the ground-state energy per particle $\varepsilon^{(\Omega)}_{\phi,0}$ (or the free energy per particle $f^{(\Omega)}_{\phi,\beta}$, for $T > 0$)

$$\varepsilon^{(\Omega)}_{\phi,0} - \varepsilon^{(0)}_{\phi,0} \equiv \lim_{V\to 0}\lim_{N\to\infty}\frac{E^{(\Omega)}_{\phi,0} - E^{(0)}_{\phi,0}}{N} > 0, \quad (22)$$

$$f^{(\Omega)}_{\phi,\beta} - f^{(0)}_{\phi,\beta} \equiv \lim_{V\to 0}\lim_{N\to\infty}\frac{\mathcal{F}^{(\Omega)}_{\phi,\beta} - \mathcal{F}^{(0)}_{\phi,\beta}}{N} > 0, \quad (23)$$

respectively. This completes the proof of our no-no theorem, prohibiting the existence of Wilczek's spontaneously rotating quantum time crystals [30].

We can also obtain interesting lower and upper bounds for the moment of inertia. The lower bound generalizes to the case of finite temperature and broken time-reversal

invariance a result given earlier by Leggett [31], and reads [28]

$$\tilde{\mathcal{I}}_{\phi,\beta} \geq \sum_i m_i R^2 \left[ 1 - \left( \left\langle \rho_{\phi,\beta}^{(i)} \right\rangle_\circ \left\langle \frac{1}{\rho_{\phi,\beta}^{(i)}} \right\rangle_\circ \right)^{-1} \right]$$
$$+ \beta \left( \langle \hat{\Lambda}_\phi^2 \rangle_{\phi,\beta} - \langle \hat{\Lambda}_\phi \rangle_{\phi,\beta}^2 \right) > 0, \qquad (24)$$

where $\langle \cdots \rangle_\circ$ indicates the average over the ring circumference, and $\rho_{\phi,\beta}^{(i)}$ is the equilibrium density for particle $i$ at flux $\phi$ and inverse temperature $\beta$. The second term, in the above result, vanishes for a system with time-reversal invariance, or at zero temperature, and did not appear in the original Leggett inequality [31]. From the Cauchy-Schwarz inequality, $\langle \rho^{(i)} \rangle_\circ \langle \frac{1}{\rho^{(i)}} \rangle_\circ \geq 1$, with equality if and only if the density $\rho^{(i)}$ is uniform, we confirm that $\tilde{\mathcal{I}}_{\phi,\beta} > 0$ for a system with broken rotational invariance. The remarkable feature of Leggett's inequality is that (at least at $T = 0$, where the second term vanishes) it allows us to obtain a lower bound for the moment of inertia in terms of the density distribution only. The upper bound reads [28]

$$\tilde{\mathcal{I}}_{\phi,\beta} \leq \beta \left( \langle \hat{L}_z^2 \rangle_{\phi,\beta} - \langle \hat{L}_z \rangle_{\phi,\beta}^2 \right), \qquad (25)$$

where the equality holds in the classical (or high temperature) limit.

Finally, we briefly comment on the proposal [3] of testing Wilczek's concept by using a Wigner crystal made of 100 $^9$Be$^+$ ions in a toroidal trap with a diameter of 100 $\mu$m, threaded by a magnetic flux. Of course, for this system, the above no-go theorem prohibits any time-crystal-like spontaneous rotation. Furthermore, for this Wigner crystal in the strong-coupling regime, simple considerations indicate that the ions are strongly localized around their classical equilibrium positions, with a Gaussian density distribution of width $w$ given by $\frac{w}{d} \propto \left( \frac{m_e}{M} \frac{a_B}{d} \right)^{1/4}$, where $d$ is the Wigner-crystal lattice parameter, $a_B$ is the Bohr radius, $m_e$ the electron mass, and $M$ the ion mass. This yields $w/d \simeq 10^{-2}$, and using Leggett's inequality, Eq. (24), we can conclude that the quantum correction to the moment of inertia is completely negligible and that this system behaves classically with respect to its rotational dynamics, in sharp contrast with the claims of the authors of Ref. [3], but in full agreement with Kohn's theory [14].

The impossibility of spontaneous ground-state rotation is nicely explained and illustrated by a simple, physically transparent model proposed by Nozières [32].

I warmly thank Philippe Nozières for numerous illuminative discussions, as well as Andres Cano for helpful comments and suggestions.

**Rotation and moment of inertia at finite temperature**

For a rotating system in thermal equilibrium at non-zero temperature, the population of the various energy levels will be determined by their energies as seen from the rotating frame, since it is in this frame that the system experiences a time-independent Hamiltonian. Thus the free energy, in the rotating frame, $\mathcal{F}'^{(\Omega)}_{\boldsymbol{\phi},\beta}$ is given by

$$e^{-\beta\mathcal{F}'^{(\Omega)}_{\boldsymbol{\phi},\beta}} \equiv \mathrm{Tr}\left(e^{-\beta\left(\hat{H}^{(0)}_{\boldsymbol{\phi}}-\Omega\hat{L}_z\right)}\right) = \sum_n e^{-\beta E'^{(\Omega)}_{\boldsymbol{\phi},n}}. \tag{S1}$$

In order to obtain the expression of the corresponding free energy in the static frame, $\mathcal{F}^{(\Omega)}_{\boldsymbol{\phi},\beta}$, we note that the energies of the levels have to be transformed from the rotating to the static frame, whereas their populations remain the same as in the rotating frame. This yields

$$\mathcal{F}^{(\Omega)}_{\boldsymbol{\phi},\beta} = \mathcal{F}'^{(\Omega)}_{\boldsymbol{\phi},\beta} + \sum_n \left[e^{-\beta(E'^{(\Omega)}_{\boldsymbol{\phi},n}-\mathcal{F}'^{(\Omega)}_{\boldsymbol{\phi},\beta})}\left(E^{(\Omega)}_{\boldsymbol{\phi},n} - E'^{(\Omega)}_{\boldsymbol{\phi},n}\right)\right]. \tag{S2}$$

From Eqs. (4, S1), we get

$$\mathcal{F}'^{(\Omega)}_{\boldsymbol{\phi},\beta} = \mathcal{F}^{(0)}_{\boldsymbol{\phi}+\Omega\boldsymbol{\mu},\beta} - \hbar\Omega\Phi_{\mathrm{tot}} - \frac{I_{\mathrm{cl}}\Omega^2}{2}. \tag{S3}$$

Together with Eq. (4), this gives

$$E'^{(\Omega)}_{\boldsymbol{\phi},n} - \mathcal{F}'^{(\Omega)}_{\boldsymbol{\phi},\beta} = E^{(0)}_{\boldsymbol{\phi}+\Omega\boldsymbol{\mu},n} - \mathcal{F}^{(0)}_{\boldsymbol{\phi}+\Omega\boldsymbol{\mu},\beta}. \tag{S4}$$

From Eqs. (4, 8, 11), we get

$$E^{(\Omega)}_{\boldsymbol{\phi},n} - E'^{(\Omega)}_{\boldsymbol{\phi},n} = \Omega L_{\boldsymbol{\phi}+\Omega\boldsymbol{\mu},n}. \tag{S5}$$

Combining Eqs. (S2, S3, S4, S5), we then obtain

$$\mathcal{F}^{(\Omega)}_{\boldsymbol{\phi},\beta} = \mathcal{F}^{(0)}_{\boldsymbol{\phi}+\Omega\boldsymbol{\mu},\beta} - \Omega\frac{d\mathcal{F}^{(0)}_{\boldsymbol{\phi}+\Omega\boldsymbol{\mu},\beta}}{d\Omega} + \frac{I_{\mathrm{cl}}\Omega^2}{2}. \tag{S6}$$

Introducing

$$\mathcal{L}_{\boldsymbol{\phi}+\Omega\boldsymbol{\mu},\beta} \equiv -\frac{d\mathcal{F}'^{(\Omega)}_{\boldsymbol{\phi},\beta}}{d\Omega} \tag{S7}$$

$$= \hbar\Phi_{\mathrm{tot}} + I_{\mathrm{cl}}\Omega - \frac{d\mathcal{F}^{(0)}_{\boldsymbol{\phi}+\Omega\boldsymbol{\mu},\beta}}{d\Omega}, \tag{S8}$$

and

$$\tilde{\mathcal{I}}_{\boldsymbol{\phi}+\Omega\boldsymbol{\mu},\beta} \equiv \frac{d\mathcal{L}_{\boldsymbol{\phi}+\Omega\boldsymbol{\mu},\beta}}{d\Omega}, \tag{S9}$$

we obtain

$$\frac{d\mathcal{F}^{(\Omega)}_{\boldsymbol{\phi},\beta}}{d\Omega} = \Omega\tilde{\mathcal{I}}_{\boldsymbol{\phi}+\Omega\boldsymbol{\mu},\beta}, \tag{S10}$$

and eventually

$$\mathcal{F}'^{(\Omega)}_{\boldsymbol{\phi},\beta} = \mathcal{F}^{(0)}_{\boldsymbol{\phi},\beta} - \Omega \mathcal{L}_{\boldsymbol{\phi},\beta} - \frac{\tilde{\mathcal{I}}_{\boldsymbol{\phi},\beta}\Omega^2}{2} + \mathcal{O}(\Omega^3), \tag{S11}$$

$$\mathcal{F}^{(\Omega)}_{\boldsymbol{\phi},\beta} = \mathcal{F}^{(0)}_{\boldsymbol{\phi},\beta} + \int_0^\Omega d\Omega'\, \Omega' \tilde{\mathcal{I}}_{\boldsymbol{\phi}+\Omega'\boldsymbol{\mu},\beta} \tag{S12}$$

$$= \mathcal{F}^{(0)}_{\boldsymbol{\phi},\beta} + \frac{\tilde{\mathcal{I}}_{\boldsymbol{\phi},\beta}\Omega^2}{2} + \mathcal{O}(\Omega^3). \tag{S13}$$

From Eqs. (11, S8), we obtain the expression of the angular momentum in rotational thermodynamic equilibrium:

$$\mathcal{L}_{\boldsymbol{\phi},\beta} = \sum_n \left( e^{-\beta(E^{(0)}_{\boldsymbol{\phi},n} - \mathcal{F}^{(0)}_{\boldsymbol{\phi},\beta})} L_{\boldsymbol{\phi},n} \right) \tag{S14}$$

$$= \langle \hat{L}_z \rangle_{\boldsymbol{\phi},\beta}, \tag{S15}$$

where

$$\langle \hat{A} \rangle_{\boldsymbol{\phi},\beta} \equiv \mathrm{Tr}\left[ e^{-\beta(\hat{H}^{(0)}_{\boldsymbol{\phi}} - \mathcal{F}^{(0)}_{\boldsymbol{\phi},\beta})} \hat{A} \right]. \tag{S16}$$

For the moment of inertia in thermodynamic equilibrium, $\tilde{\mathcal{I}}_{\boldsymbol{\phi},\beta}$, using Eqs. (S9, S14), we obtain

$$\tilde{\mathcal{I}}_{\boldsymbol{\phi},\beta} = \mathcal{I}_{\boldsymbol{\phi},\beta} + \beta \left( \langle \hat{\Lambda}^2_{\boldsymbol{\phi}} \rangle_{\boldsymbol{\phi},\beta} - \langle \hat{\Lambda}_{\boldsymbol{\phi}} \rangle^2_{\boldsymbol{\phi},\beta} \right), \tag{S17}$$

with

$$\mathcal{I}_{\boldsymbol{\phi},\beta} \equiv \sum_{n \neq m} \left( \frac{e^{-\beta E^{(0)}_{\boldsymbol{\phi},n}} - e^{-\beta E^{(0)}_{\boldsymbol{\phi},m}}}{e^{-\beta \mathcal{F}^{(0)}_{\boldsymbol{\phi},\beta}}\left( E^{(0)}_{\boldsymbol{\phi},m} - E^{(0)}_{\boldsymbol{\phi},n} \right)} |L_{\boldsymbol{\phi},n,m}|^2 \right), \tag{S18}$$

$$\hat{\Lambda}_{\boldsymbol{\phi}} \equiv \sum_n |\varphi^0_{\boldsymbol{\phi},n}\rangle L_{\boldsymbol{\phi},n} \langle \varphi^0_{\boldsymbol{\phi},n}|. \tag{S19}$$

**Generalized Leggett inequality**

Leggett [1] used the Rayleigh-Ritz variational principle to generate an upper bound of the ground state energy $E^{(0)}_{\boldsymbol{\phi}+\Omega\boldsymbol{\mu},0}$. For the finite temperature case, we can get an upper bound for the free energy $\mathcal{F}_{\boldsymbol{\phi}+\Omega\boldsymbol{\mu},\beta}$ (since all quantities here and below refer exclusively to the static frame, we shall omit the upper indices) by using the Gibbs-Bogoliubov variational principle [2]:

$$\mathcal{F}_{\boldsymbol{\phi}+\Omega\boldsymbol{\mu},\beta} \leq \mathcal{F}^{\mathrm{trial}}_{\boldsymbol{\phi}+\Omega\boldsymbol{\mu},\beta} + \langle \hat{H}_{\boldsymbol{\phi}+\Omega\boldsymbol{\mu}} - \hat{H}^{\mathrm{trial}}_{\boldsymbol{\phi}+\Omega\boldsymbol{\mu}} \rangle_{\mathrm{trial}}, \tag{S20}$$

with

$$e^{-\beta \mathcal{F}^{\mathrm{trial}}_{\boldsymbol{\phi}+\Omega\boldsymbol{\mu},\beta}} \equiv \mathrm{Tr}\left( e^{-\beta \hat{H}^{\mathrm{trial}}_{\boldsymbol{\phi}+\Omega\boldsymbol{\mu}}} \right), \tag{S21}$$

$$\langle \hat{A} \rangle_{\mathrm{trial}} \equiv \mathrm{Tr}\left( e^{-\beta(\hat{H}^{\mathrm{trial}}_{\boldsymbol{\phi}+\Omega\boldsymbol{\mu}} - \mathcal{F}^{\mathrm{trial}}_{\boldsymbol{\phi}+\Omega\boldsymbol{\mu},\beta})} \hat{A} \right). \tag{S22}$$

Generalizing Leggett's ansatz to the finite temperature case, we chose the trial Hamiltonian $\hat{H}^{\mathrm{trial}}_{\boldsymbol{\phi}+\Omega\boldsymbol{\mu}}$ as follows. Let

$$\varphi_{\boldsymbol{\phi},n}(\boldsymbol{\theta}) \equiv \sqrt{\rho_{\boldsymbol{\phi},n}(\boldsymbol{\theta})}\, e^{i\alpha_{\boldsymbol{\phi},n}(\boldsymbol{\theta})} \tag{S23}$$

be the exact ground-state many-body wave-functions for fluxes $\boldsymbol{\phi}$. We introduce the trial wave-functions for fluxes $\boldsymbol{\phi} + \Omega\boldsymbol{\mu}$

$$\varphi^{\mathrm{trial}}_{\boldsymbol{\phi}+\Omega\boldsymbol{\mu},n}(\boldsymbol{\theta}) \equiv \varphi_{\boldsymbol{\phi},n}(\boldsymbol{\theta})\, e^{i\sum_i \beta_i(\theta_i)}. \tag{S24}$$

where the trial phase functions $\boldsymbol{\beta} \equiv (\beta_1, \beta_2, \ldots, \beta_N)$ satisfy the boundary conditions

$$\beta_i(\theta_i + 2\pi) = \beta_i(\theta_i). \tag{S25}$$

For discernable particles, the phase functions $\beta_i$ may be generally different. For identical particles obeying Bose-Einstein of Fermi-Dirac statistics, we impose upon the phase functions to be identical, so that the phase factor is symmetrical under permutation, and the overall symmetry of the trial wave-function with respect to permutations is the same as for the reference wave-functions $\varphi_{\phi,n}(\boldsymbol{\theta})$. The corresponding trial energies are

$$E^{\text{trial}}_{\boldsymbol{\phi}+\Omega\boldsymbol{\mu},n} \equiv \langle \varphi^{\text{trial}}_{\boldsymbol{\phi}+\Omega\boldsymbol{\mu},n} | \hat{H}_{\boldsymbol{\phi}+\Omega\boldsymbol{\mu}} | \varphi^{\text{trial}}_{\boldsymbol{\phi}+\Omega\boldsymbol{\mu},n} \rangle, \tag{S26}$$

and we construct the trial Hamiltonian as

$$\hat{H}^{\text{trial}}_{\boldsymbol{\phi}+\Omega\boldsymbol{\mu}} \equiv \sum_n |\varphi^{\text{trial}}_{\boldsymbol{\phi}+\Omega\boldsymbol{\mu},n}\rangle \, E^{\text{trial}}_{\boldsymbol{\phi}+\Omega\boldsymbol{\mu},n} \, \langle \varphi^{\text{trial}}_{\boldsymbol{\phi}+\Omega\boldsymbol{\mu},n} |. \tag{S27}$$

For this trial Hamiltonian, the second term in Eq. (S20) vanishes, and the Gibbs-Bogoliubov inequality reads

$$\mathcal{F}_{\boldsymbol{\phi}+\Omega\boldsymbol{\mu},\beta} \leq \mathcal{F}^{\text{trial}}_{\boldsymbol{\phi}+\Omega\boldsymbol{\mu},\beta}, \tag{S28}$$

with

$$e^{-\beta \mathcal{F}^{\text{trial}}_{\boldsymbol{\phi}+\Omega\boldsymbol{\mu},\beta}} = \sum_n e^{-\beta E^{\text{trial}}_{\boldsymbol{\phi}+\Omega\boldsymbol{\mu},n}}. \tag{S29}$$

With the above ansatz, the trial energy eigenvalues are

$$E^{\text{trial}}_{\boldsymbol{\phi}+\Omega\boldsymbol{\mu},n} = \int \prod_j d\theta_j \sum_i \frac{\hbar}{2\mu_i} \left| (-i\partial_{\theta_i} - \phi_i - \Omega\mu_i) \varphi^{\text{trial}}_{\boldsymbol{\phi}+\Omega\boldsymbol{\mu},n} \right|^2 + V[\rho_{\boldsymbol{\phi},n}] + U[\rho_{\boldsymbol{\phi},n}] \tag{S30}$$

$$= E_{\boldsymbol{\phi},n} + \sum_i \frac{\hbar}{\mu_i} \int d\theta_i \, (\partial_{\theta_i}\beta_i - \Omega\mu_i) \int \prod_{j(\neq i)} d\theta_j \, \rho_{\boldsymbol{\phi},n} \, (\partial_{\theta_i}\alpha_{\boldsymbol{\phi},n} - \phi_i)$$

$$+ \sum_i \frac{\hbar}{2\mu_i} \int d\theta_i \, \rho^{(i)}_{\boldsymbol{\phi},n} \, (\partial_{\theta_i}\beta_i - \Omega\mu_i)^2. \tag{S31}$$

In the above equation,

$$\rho^{(i)}_{\boldsymbol{\phi},n}(\theta_i) \equiv \int \prod_{j(\neq i)} d\theta_j \, \rho_{\boldsymbol{\phi},n}(\boldsymbol{\theta}) \tag{S32}$$

is the probability density for particle $i$. Noting that

$$\frac{\hbar}{\mu_i} \int \prod_{j(\neq i)} d\theta_j \, \rho_{\boldsymbol{\phi},n} \, (\partial_{\theta_i}\alpha_{\boldsymbol{\phi},n} - \phi_i) \tag{S33}$$

expresses the current density for particle $i$ in the stationary state $\varphi_{\boldsymbol{\phi},n}(\boldsymbol{\theta})$, and is therefore independent of $\theta_i$, we get

$$E^{\text{trial}}_{\boldsymbol{\phi}+\Omega\boldsymbol{\mu},n} = E_{\boldsymbol{\phi},n} - \Omega \langle \hat{L}_z \rangle_{\boldsymbol{\phi},n} + \sum_i \frac{\hbar}{2\mu_i} \int d\theta_i \, \rho^{(i)}_{\boldsymbol{\phi},n} \, (\partial_{\theta_i}\beta_i - \Omega\mu_i)^2, \tag{S34}$$

where

$$\langle \hat{L}_z \rangle_{\boldsymbol{\phi},n} \equiv \langle \varphi_{\boldsymbol{\phi},n} | \hat{L}_z | \varphi_{\boldsymbol{\phi},n} \rangle \tag{S35}$$

is the angular momentum for the state $\varphi_{\boldsymbol{\phi},n}(\boldsymbol{\theta})$. Anticipating the fact (to be proven below) that the last term in Eq. (S34) is proportional to $\Omega^2$, we write it as

$$\sum_i \frac{\hbar}{2\mu_i} \int d\theta_i \, \rho^{(i)}_{\boldsymbol{\phi},n} \, (\partial_{\theta_i}\beta_i - \Omega\mu_i)^2 \equiv A_n \frac{\Omega^2}{2}, \tag{S36}$$

so that

$$\frac{dE^{\text{trial}}_{\boldsymbol{\phi}+\Omega\boldsymbol{\mu},n}}{d\Omega} = -\langle \hat{L}_z \rangle_{\boldsymbol{\phi},n} + \Omega A_n, \tag{S37}$$

$$\frac{d^2 E^{\text{trial}}_{\boldsymbol{\phi}+\Omega\boldsymbol{\mu},n}}{d\Omega^2} = A_n. \tag{S38}$$

From Eq. (S28), we have

$$\frac{d\mathcal{F}^{\text{trial}}_{\phi+\Omega\mu,\beta}}{d\Omega} = \sum_n e^{-\beta(E^{\text{trial}}_{\phi+\Omega\mu,n}-\mathcal{F}^{\text{trial}}_{\phi+\Omega\mu,\beta})} \frac{dE^{\text{trial}}_{\phi+\Omega\mu,n}}{d\Omega}, \tag{S39}$$

$$\frac{d^2\mathcal{F}^{\text{trial}}_{\phi+\Omega\mu,\beta}}{d\Omega^2} = \sum_n e^{-\beta(E^{\text{trial}}_{\phi+\Omega\mu,n}-\mathcal{F}^{\text{trial}}_{\phi+\Omega\mu,\beta})} \left[\frac{d^2 E^{\text{trial}}_{\phi+\Omega\mu,n}}{d\Omega^2} - \beta \frac{dE^{\text{trial}}_{\phi+\Omega\mu,n}}{d\Omega} \frac{d\left(E^{\text{trial}}_{\phi+\Omega\mu,n}-\mathcal{F}^{\text{trial}}_{\phi+\Omega\mu,\beta}\right)}{d\Omega}\right]. \tag{S40}$$

This yields

$$\mathcal{F}^{\text{trial}}_{\phi+\Omega\mu,\beta}\Big|_{\Omega=0} = \mathcal{F}_{\phi,\beta}, \tag{S41}$$

$$\frac{d\mathcal{F}^{\text{trial}}_{\phi+\Omega\mu,\beta}}{d\Omega}\bigg|_{\Omega=0} = -\langle \hat{L}_z \rangle_{\phi,\beta} = \frac{d\mathcal{F}_{\phi+\Omega\mu,\beta}}{d\Omega}\bigg|_{\Omega=0}, \tag{S42}$$

$$\frac{d^2\mathcal{F}^{\text{trial}}_{\phi+\Omega\mu,\beta}}{d\Omega^2}\bigg|_{\Omega=0} = \frac{1}{\Omega^2}\sum_i \frac{\hbar}{2\mu_i}\int d\theta_i \rho^{(i)}_{\phi,\beta}(\partial_{\theta_i}\beta_i - \Omega\mu_i)^2 - \beta\left(\langle \hat{\Lambda}_\phi^2\rangle_{\phi,\beta} - \langle \hat{\Lambda}_\phi\rangle_{\phi,\beta}^2\right), \tag{S43}$$

where

$$\rho^{(i)}_{\phi,\beta} \equiv \sum_n e^{-\beta(E_{\phi,n}-\mathcal{F}_{\phi,\beta})}\rho^{(i)}_{\phi,n} \tag{S44}$$

is the probability density for particle $i$ in thermal equilibrium. This implies that

$$\frac{d^2\mathcal{F}_{\phi+\Omega\mu,\beta}}{d\Omega^2}\bigg|_{\Omega=0} \leq \frac{1}{\Omega^2}\sum_i \frac{\hbar}{2\mu_i}\int d\theta_i \rho^{(i)}_{\phi,\beta}(\partial_{\theta_i}\beta_i - \Omega\mu_i)^2 - \beta\left(\langle \hat{\Lambda}_\phi^2\rangle_{\phi,\beta} - \langle \hat{\Lambda}_\phi\rangle_{\phi,\beta}^2\right). \tag{S45}$$

The minimization of the first term on the right-hand side of the above equation using the Euler-Lagrange method yields

$$\frac{d}{d\theta_i}\left[\rho^{(i)}_{\phi,\beta}(\partial_{\theta_i}\beta_i - \Omega\mu_i)\right] = 0, \tag{S46}$$

which gives

$$(\partial_{\theta_i}\beta_i - \Omega\mu_i) = -\Omega\mu_i \left(\rho^{(i)}_{\phi,\beta}\left\langle\frac{1}{\rho^{(i)}_{\phi,\beta}}\right\rangle_\circ\right)^{-1} \tag{S47}$$

and justifies the statement expressed by Eq. (S36). Inserting the latter result in Eq. (S45), we eventually obtain the generalized Leggett inequality

$$\tilde{\mathcal{I}}_{\phi,\beta} \geq \sum_i m_i R^2 \left[1 - \left(\left\langle\rho^{(i)}_{\phi,\beta}\right\rangle_\circ \left\langle\frac{1}{\rho^{(i)}_{\phi,\beta}}\right\rangle_\circ\right)^{-1}\right] + \beta\left(\langle \hat{\Lambda}_\phi^2\rangle_{\phi,\beta} - \langle \hat{\Lambda}_\phi\rangle_{\phi,\beta}^2\right). \tag{S48}$$

**Upper bound for the moment of inertia**

We use the Golden-Thompson inequality [3, 4]

$$\text{Tr}\left(e^{A+B}\right) \leq \text{Tr}\left(e^A e^B\right), \tag{S49}$$

for any finite dimensional Hermitian matrices $A$ and $B$, to obtain a lower bound for the free energy $\mathcal{F}'^{(\Omega)}_{\phi,\beta}$, given by

$$e^{-\beta \mathcal{F}'^{(\Omega)}_{\phi,\beta}} = \text{Tr}\left(e^{-\beta(\hat{H}_\phi - \Omega\hat{L}_z)}\right) \tag{S50}$$

$$\leq \text{Tr}\left(e^{-\beta\hat{H}_\phi}e^{\beta\Omega\hat{L}_z}\right). \tag{S51}$$

Expanding both sides of this expression to second order in $\Omega$, and proceeding as above, we eventually obtain the upper bound for the moment of inertia:

$$\tilde{\mathcal{I}}_{\phi,\beta} \leq \beta \left( \langle \hat{L}_z^2 \rangle_{\phi,\beta} - \langle \hat{L}_z \rangle_{\phi,\beta}^2 \right). \tag{S52}$$

The inequality becomes an equality when the matrix elements of commutator $\left[ \hat{L}_z, \hat{H}_\phi \right]$ are much smaller than $\beta^{-1}$, which corresponds to the classical limit, and for which it yields the classical moment of inertia $I_{\rm cl}$ [5].

---